\documentclass[aps,prc,preprintnumbers,epsf]{revtex4}
\newcommand{\gtap}{\;{\raise.3ex\hbox{$>$\kern-.75em\lower1ex\hbox{$\sim$}}}\;}
\newcommand{\ltap}{\;{\raise.3ex\hbox{$<$\kern-.75em\lower1ex\hbox{$\sim$}}}\;}
\usepackage[dvips]{graphicx}
\newcommand{\bea}{\begin{eqnarray}}
\newcommand{\eea}{\end{eqnarray}}

\begin{document}
\preprint{arXiv: 0901.2766[nucl-th]}
\title{Nonperturbative $^3S_1$$-$$^3D_1$ $NN$ scattering in pionless EFT}
\author{Ji-Feng Yang}
\address{Department of Physics, East China Normal University,
Shanghai, 200062, China}
\date{January 15, 2009}
\begin{abstract}
The rigorous on-shell $T$-matrices for $NN$ scattering in the
coupled channels $^3S_1$$-$$^3D_1$ are briefly presented in the
context of EFT($\not\!\!\pi$) with the contact potentials truncated
at order $\Delta=4$. The nonperturbative features of renormalization
are highlighted and elaborated. A simple scenario for EFT power
counting's and renormalization prescriptions is also presented with
its consequences being roughly analyzed and discussed.
\end{abstract}\maketitle{\em
Introduction.} Since Weinberg's seminal proposal\cite{WeinEFT},
there have been enormous progresses in the effective field theory
(EFT) approach to nucleon systems\cite{Epelrev}, pointing towards a
more field theoretical treatment of the nuclear forces basing on
quantum chromodynamics. Meanwhile, some new theoretical issues also
arise. The most intriguing one is to construct a unified framework
for renormalization of such EFT and power counting in
nonperturbative regime. There have appeared several new power
counting schemes\cite{KSW,BBSvK,NTvK}, but the strong interplay
between nonperturbative renormalization and power counting makes a
completely satisfactory answer still out of reach\cite{kaplan}, for
recent debates, see\cite{NTvK,EpelMeis,Review}. In our view, it is
desirable and also easy to explore the issue with simple but
rigorous solutions where the main difficulties may become
transparent. Such an attempt has been made in $^1S_0$ channel in
Ref.\cite{C71}, offering an alternative perspective to the
intriguing issue. Recently\cite{scenario}, we argued that the main
nonperturbative features revealed in Ref.\cite{C71} should be
universal in all channels. In this short report, we briefly present
the rigorous solutions for the coupled channels $^3S_1$$-$$^3D_1$
with contact potentials that could be very easily obtained. As will
be clear below, the main characteristics of nonperturbative
renormalization highlighted in the lowest uncoupled channel
$^1S_0$\cite{C71} and recently argued to be universal in
\cite{scenario} is indeed true in these coupled channels. We will
also briefly present a simple scenario ensued by a rough analysis of
some possible predictions of physical behaviors. The details along
with further analysis and comparisons across different EFT power
counting schemes and renormalization prescriptions will be given in
a forthcoming report.

{\em Rigorous solutions.} Let us start with the Lippmann-Schwinger
equations (LSE) for coupled channels $^3S_1$$-$$^3D_1$,
\bea\label{LSE} {\bf T}(q,q^{\prime}; E)={\bf V}(q,q^{\prime})+\int
\displaystyle\frac{d^3k}{(2\pi)^3} \frac{{\bf V}(q,k){\bf
T}(k,q^{\prime}; E)}{E-k^2/M+i\epsilon},\ \ \ {\bf{V}}\equiv
\left(\begin{array}{cc} V_{ss}& V_{sd}\\V_{ds} &V_{dd}\\
\end{array}\right),\ {\bf{T}}\equiv \left(\begin{array}{cc}
T_{ss} & T_{sd}\\T_{ds} &T_{dd}\\ \end{array}\right),\eea with $E$
denoting the center of mass energy, $M$ the nucleon mass,
$q,q^{\prime}$ the off-shell external momenta. For the contact
potential truncated at a finite chiral order $\Delta$, one could
introduce a column vector $U(q)$, whose transpose is defined as
$U^T(q)\equiv (1,q^2,q^4,\cdots)$, to 'factorize' $V_{xy}$ and
$T_{xy}$ as: $V_{xy}=U^T\lambda_{xy}U,\ \ T_{xy}=U^T\tau_{xy}U$,
where $\lambda$ and $\tau$ are matrices\cite{PBC}. At $\Delta=4$, we
have, $U^T(q)=(1,q^2,q^4)$ and\bea \lambda_{ss}\equiv \left(
\begin{array}{ccc}  C_{0;ss}& C_{2;ss} &C_{4;ss} \\
  C_{2;ss} & \tilde{C}_{4;ss} &0 \\ C_{4;ss} &0 & 0 \\
\end{array}\right);\ \lambda_{sd}=\lambda_{ds}^T\equiv \left(%
\begin{array}{ccc} 0 & 0 &0 \\ C_{2;sd} & \tilde{C}_{4;sd} &0 \\
C_{4;sd} &0 & 0 \\ \end{array} \right); \ \lambda_{dd}\equiv \left(%
\begin{array}{ccc}  0 & 0 &0\\ 0 & \tilde{C}_{4;dd} &0 \\
0 &0 & 0 \\ \end{array} \right).\nonumber\eea Conventionally, the
contact EFT couplings $[C_{\Delta;\ldots}]$ scale like
$C_{\Delta;\ldots}/C_{0;ss}\sim \Lambda^{-\Delta} $ with $\Lambda$
being the upper scale for EFT.

Using the above 'factorization' trick, Eq.(\ref{LSE}) become
algebraic, \bea \label{algebrLSE}
&&\underline{\tau}=\underline{\lambda} +\underline{\lambda}
\underline{{\mathcal{I}}}(E)\underline{\tau}, \ \ \ \
\underline{\lambda}\equiv \left(\begin{array}{cc}
\lambda_{ss}& \lambda_{sd}\\\lambda_{ds} &\lambda_{dd}\\
\end{array}\right),\ \ \underline{\tau}\equiv
\left(\begin{array}{cc}\tau_{ss} &\tau_{sd}\\
\tau_{ds} &\tau_{dd}\\\end{array}\right),\ \
\underline{{\mathcal{I}}}(E)\equiv \left(\begin{array}{cc}
{\mathcal{I}}(E) & {\bf 0} \\{\bf 0} &{\mathcal{I}}(E)\\
\end{array}\right),\eea where the matrix
$\label{I}{\mathcal{I}}(E)\equiv\int \frac{d^3k}{(2\pi)^3}
\frac{U(k)U^T(k)}{E-k^2/M+i\epsilon}$ comprises of the divergent
integrals arising from the convolution. A general element of this
matrix could then be parametrized as follows
($p\equiv\sqrt{ME}$):\bea \int \frac{d^3k}{(2\pi)^3}
\frac{k^{2n}}{E-k^2/M+i\epsilon}= \sum_{m=1}^{n}J_{2m+1}p^{2(n-m)}-
{\mathcal{I}}_0p^{2n}, \ \mathcal{I}_0\equiv
J_0+\frac{iMp}{4\pi},\eea where $\{J_n\}$ ($n=0,3,5,\cdots$) are
real constants and preliminarily regularization and/or
renormalization prescription dependent. ${\mathcal{I}}(E)$ could be
further casted into the following form \bea
\label{I-short}{\mathcal{I}}(E)=-\mathcal{I}_0 U(p)U^T(p) +J_3\Delta
U_1+J_5\Delta U_2+\cdots\eea with $\Delta
U_1\equiv\frac{1}{p^2}\int^{p^2}_0 dt\frac{d[U(t)U^T(t)]}{dt},\
\Delta U_{n+1}\equiv\frac{1}{p^2}\int^{p^2}_0 dt \frac{d[\Delta
U_n(t)]}{dt},\ n=1,2,\cdots$. Now the solutions for the
$\tau_{xy}$'s are easy to find from the solution \bea
\underline{\tau}=\left(1-\underline{\lambda}
\underline{{\mathcal{I}}}(E)\right)^{-1}\underline{\lambda}.\eea For
example, $\tau_{ss}(E)=(1-\tilde{\lambda}_{ss}
{\mathcal{I}}(E))^{-1} \tilde{\lambda}_{ss}$, with $
\tilde{\lambda}_{ss}\equiv\lambda_{ss}+\lambda_{sd}{\mathcal{I}}(E)
(1-\lambda_{dd}{\mathcal{I}}(E) )^{-1}\lambda_{ds}$.

Then, the on-shell $T$-matrices could be readily obtained as
$T_{xy}=U^T(p)\tau_{xy}U(p)$ with $p=\sqrt{ME}$. At $\Delta=4$, we
find ($\mathcal{I}_0\equiv J_0+\frac{iMp}{4\pi}$):\bea
\label{invTss4} &&\frac{1}{T_{ss}(p)}=\mathcal{I}_0+
\frac{{\mathcal{N}}_0
+\mathcal{I}_0{\mathcal{N}}_1p^4}{{\mathcal{D}}_0
+\mathcal{I}_0{\mathcal{D}}_1p^4}, \ \ \ \ \ \ \
\frac{1}{T_{dd}(p)}=\mathcal{I}_0+ \frac{{\mathcal{N}}_0
+\mathcal{I}_0 {\mathcal{D}}_0}{[{\mathcal{N}}_1
+\mathcal{I}_0{\mathcal{D}}_1]p^4},\\
\label{invTsd4}&&\frac{1}{T_{sd}(p)}=\frac{1}{T_{ds}(p)}=
\frac{{\mathcal{N}}_0+\mathcal{I}_0[{\mathcal{D}}_0
+{\mathcal{N}}_1p^4]+
\mathcal{I}_0^2{\mathcal{D}}_1p^4}{{\mathcal{D}}_{sd}p^2},\eea
with\bea \label{keydsd}&& {\mathcal{D}}_{sd}^2+{\mathcal{D}}_1
{\mathcal{N}}_0={\mathcal{N}}_1{\mathcal{D}}_0. \eea Here
$[{\mathcal{N}},{\mathcal{D}}]$ are lengthy polynomials in terms of
real parameters: $[C_{\cdots}]$, $[J_n](n\neq0)$ and $p$, they are
all $\mathcal{I}_0$-independent:
$\partial_{\mathcal{I}_0}[\mathcal{N},\mathcal{D}]=0$. The details
will be given in a forthcoming report\cite{forth}. At order
$\Delta=2$, the expressions become simpler:\bea
\label{invTss2}&&\frac{1}{T_{ss}(p)}=\mathcal{I}_0+
\frac{{\mathcal{N}}_0}{{\mathcal{D}}_0
+\mathcal{I}_0{\mathcal{D}}_1p^4}, \ \ \ \ \ \ \
\frac{1}{T_{dd}(p)}=\mathcal{I}_0+
\frac{{\mathcal{N}}_0+\mathcal{I}_0{\mathcal{D}}_0}
{\mathcal{I}_0{\mathcal{D}}_1p^4},\\
\label{invTsd2}&&\frac{1}{T_{sd}(p)}=\frac{1}{T_{ds}(p)}=
\frac{{\mathcal{N}}_0+\mathcal{I}_0{\mathcal{D}}_0+
\mathcal{I}_0^2{\mathcal{D}}_1p^4}{{\mathcal{D}}_{sd}p^2},\eea as
$[{\mathcal{N}},{\mathcal{D}}]$ become simpler at this order:
${\mathcal{D}}_0\equiv C_{0;ss}+(C_{2;ss}^2+C_{2;sd}^2)J_5+
[2C_{2;ss}+(C_{2;sd}^2-C_{2;ss}^2)J_3]p^2,\ {\mathcal{D}}_1\equiv
-C^2_{2;sd}, \ {\mathcal{D}}_{sd}\equiv C_{2;sd}(1-C_{2;ss}J_3),\
{\mathcal{N}}_0\equiv (1-C_{2;ss}J_3)^2$ and ${\mathcal{N}}_1=0$.

Although the on-shell $T$-matrices obtained above appear more
complicated in comparison with that of uncoupled channels (e.g.,
$^1S_0$) where $1/T={\mathcal{I}_0}+N/D$, with $N,D$ being
${\mathcal{I}_0}$-independent, the inverse of $\textbf{T}$ that
assembles the four matrices $T_{xy}$ appear quite simpler. Using the
relation (\ref{keydsd}), we find,\bea {\bf
T}^{-1}={\mathcal{I}_0}{\bf I}+\frac{1}{{\mathcal{D}}_1p^4}
\left(\begin{array}{cc} {\mathcal{N}}_1p^4 ,& -{\mathcal{D}}_{sd}p^2
\\-{\mathcal{D}}_{sd}p^2, &
{\mathcal{D}}_0\\\end{array} \right), \eea with ${\bf I}$ being the
$2\times 2$ unit matrix. For each entry, that is, \bea
\label{INVTss} \left ({\bf T}^{-1}\right)_{ss}={\mathcal{I}_0}
+\frac{\mathcal{N}_1}{\mathcal{D}_1},\ \ \ \ \left({\bf
T}^{-1}\right)_{dd}={\mathcal{I}_0}
+\frac{\mathcal{D}_0}{\mathcal{D}_1p^4}, \ \ \ \ \ \left({\bf
T}^{-1}\right)_{sd}=\left({\bf
T}^{-1}\right)_{ds}=-\frac{\mathcal{D}_{sd}}{\mathcal{D}_1p^2}.\eea
Here we note that $\textbf{T}^{-1}$is also simpler as $[\mathcal{N},
\mathcal{D}]$ become simpler and $\mathcal{N}_1=0$, while at leading
order $\textbf{T}^{-1}$ is singular as $\mathcal{D}_1=0$ or
$T_{dd}=T_{sd}=0$. With this inverse form, it is trivial to verify
the on-shell unitarity:
$\textbf{T}^{-1}-(\textbf{T}^{-1})^{\dagger}=
i\frac{Mp}{2\pi}\textbf{I}$. Both this unitarity and the inverse
form of $\textbf{T}$ exhibited above can be shown to hold true at
any order within the context of contact potential, the detailed
proof will be presented in the forthcoming report\cite{forth}. For
the uncoupled channels $1/T$ is just the inverse of $T$, therefore
we conclude that: \bea\label{I_indpdt}
\partial_{{\mathcal{I}_0}}(\textbf{T}^{-1}-{\mathcal{I}_0}{\bf
I})=0 \eea for any channel. Or more precisely, the real rational
part of the inverse $\textbf{T}$ in any channel is independent of
${\mathcal{I}_0}$ within the realm of contact potential or
EFT($\not\!\pi$). As will be clear shortly, this fact is
consequential, as in the $^1S_0$ case\cite{C71}. We note in passing
that this form of $\textbf{T}^{-1}$ is consistent with the standard
parametrization of $S$-matrix\cite{Stapp} where the inverse
$\textbf{T}$ reads \bea {\bf T}^{-1}=i\frac{Mp}{4\pi}{\bf
I}+\frac{Mp}{4\pi} \left(\begin{array}{cc}
\frac{\sin(\delta_s+\delta_d)- \sin(\delta_s
-\delta_d)\cos(2\epsilon)}{\cos(\delta_s+ \delta_d) -\cos(\delta_s
-\delta_d)\cos(2\epsilon)},&
\frac{-\sin(2\epsilon)}{\cos(\delta_s+\delta_d)
-\cos(\delta_s -\delta_d)\cos(2\epsilon)}  \\
\frac{-\sin(2\epsilon)}{\cos(\delta_s+ \delta_d) -\cos(\delta_s
-\delta_d)\cos(2\epsilon)}, & \frac{\sin(\delta_s+\delta_d)-
\sin(\delta_d -\delta_s)\cos(2\epsilon)}{\cos(\delta_s+ \delta_d)
-\cos(\delta_d
-\delta_s)\cos(2\epsilon)}\\
\end{array}\right),\eea with ${\bf S}\equiv {\bf I}-i\frac{Mp}{2\pi}{\bf
T}$.

{\em Nonperturbative renormalization.} As divergences appear in both
numerators and denominators (e.g., in $\mathcal{N}_0, \mathcal{N}_1,
\mathcal{D}_0, \mathcal{D}_1$ and $\mathcal{D}_{sd}$) of the compact
$T$-matrices, they must be renormalized or rendered finite in such a
manner that the $p$-dependence of the $T$ matrices be preserved.
Examining the simple results at $\Delta=2$ given above, it is easy
to see that the perturbative counterterm algorithm could not work
here. Thus the renormalization must be done somehow
nonperturbatively. Then the compact form of $T$'s leads to
nontrivial prescription dependence. This in turn implies a strong
interplay between renormalization prescriptions and EFT power
counting\cite{C71}. Hence a consistent framework must fully
appreciate and explore this fact.

In particular, since
$\partial_{{\mathcal{I}_0}}(\textbf{T}^{-1}-{\mathcal{I}_0}{\bf
I})=0$, ${\mathcal{I}_0}$ is 'isolated' from or 'decoupled' with the
real rational parts of $\textbf{T}^{-1}$. Moreover, the functional
form of these rational parts (e.g., $\mathcal{N}_1/\mathcal{D}_1$)
could never accommodate a constant that could be absorbed into
${\mathcal{I}_0}$, unless one forces an expansion on the rational
parts that would ruin the nonperturbative status. For example, at
order $\Delta=2$, we even have $({\bf T}^{-1})_{ss}={\mathcal{I}_0}$
as $\mathcal{N}_1=0 (\mathcal{D}_{1}\neq0)$. (At leading order
$\Delta=0$, $\textbf{T}^{-1}$ is singular and meaningless as
$\mathcal{D}_{1}=0$.) Therefore, ${\mathcal{I}_0}$ must be
renormalized separately. In fact, as the $p$-dependence of
$\textbf{T}^{-1}$ is also physical, the isolated 'position' of
${\mathcal{I}_0}$ implies that it is physical and hence $J_0$ must
be physically determined. To abuse the conventional terminology,
$J_0$ is a renormalization group (RG) invariant scale\cite{C71}.
Such kind of RG invariant parameter has been predicted within the
Wilsonian approach\cite{Birse}: [[$(\hat{V}_0)^{-1}$]], which is
just $-J_0=-\text{Re}({\mathcal{I}_0})$ computed in the Wilsonian
cutoff approach. Conventionally, renormalized objects must also be
confronted with physical boundary conditions\cite{sterman}, though
it is often less prominent due to renormalizability.

Another point worth emphasis is: there are only finite many
divergences to be removed in the nonperturbative formulation at a
finite order of potential truncation, in spite that there are
formally infinite many divergences in the iterative solution of LSE.
That is, only a finite number of $[J_0, J_{n}, n=3,5,\cdots]$ are
involved in the compact form of $T$ at any truncation order. In our
view, this finiteness underlies and substantiates the tractability
of the nonperturbative renormalization of $T$ through whatever
means\cite{EPVRAM,gege}. That the nonperturbative $T$-matrices could
not be renormalized in conventional fashion does not imply that they
could not be renormalized at all. Ultimately, renormalization is to
render the objects in concern finite and comparable with
experimental data. In this connection, we feel it a simple choice to
perform the subtraction at the level of integrals somehow and then
fix the residual ambiguities with appropriate boundary
conditions\cite{gege} in a manner consistent with EFT principles.
Here, at least in the context of contact potentials or
EFT($\not\!\pi$), the divergences could be easily identified and
subtracted in the nonperturbative formulation.

{\em A simple scenario.} To perform some heuristic analysis, we need
be more specific about the power counting for both the EFT couplings
and the prescription parameters $[J_{\cdots}]$. As EFT power
counting are usually established on physical reasonings, the
simplest choice at our discretion should lie in $[J_{\cdots}]$,
leaving the EFT power counting intact. Thus we consider the
following scenario that is first considered in \cite{C71}:
\bea\label{scenario} |C_{\Delta;\cdots}|\sim
\frac{4\pi}{M}\Lambda^{-\Delta-1}, \ J_0\sim \frac{M}{4\pi}\Lambda,
\ J_{2n+1}\sim\frac{M}{4\pi}\mu^{2n+1}=
\frac{M}{4\pi}\left(\epsilon\Lambda\right)^{2n+1}, \ \
\epsilon\ll1.\eea Here $J_0$ is so chosen for the following reasons:
(1) it is a physical scale and hence it should be different from
$[J_n, n\neq0]$ which are sized as usual renormalization parameters;
(2) it could lead to an unnatural $S$-wave scattering length with
natural EFT couplings. With such a scenario, we could examine the
magnitudes of the effective range expansion (ERE) parameters for the
$^3S_1$ and $^3D_1$ channels with the $T$-matrices obtained above,
according to the following definition,\bea \label{ERT}\text{Re}\left
(-\frac{4\pi p^L}{MT_L}\right )=-\frac{1}{a}+\frac{1}{2}r_e
p^2+\sum_{k=2}^{\infty} v_k p^{2k}, \eea with the parameters $a$ and
$r_e$ being the scattering length and the effective range. In
reality these parameters (including $[v_k]$) could be extracted from
the scattering data, and imposed as boundary conditions for the
$T$-matrices. Here, they are employed to illustrate the interesting
consequences of the simple scenario and the importance of the
renormalization prescription within the context of contact
potential.

According to the scenario (\ref{scenario}), the magnitudes for the
ERE parameters in $^3S_1$ and $^3D_1$ channels at order $\Delta=4$
would be as follows: \bea
\text{Re}\left(-\frac{4\pi}{MT_{ss}}\right) &&=\Lambda\left
\{-1-\frac{1+{\mathcal{O}}(\epsilon^3)}{\tilde{c}_0}
+\frac{p^2}{\Lambda^2}{\mathcal{O}}(1+{\mathcal{O}}(\epsilon))
+\frac{p^4}{\Lambda^4}{\mathcal{O}}(1+{\mathcal{O}}(\epsilon))
+\cdots\right\}\nonumber\\
&&=\Lambda\left \{-{\mathcal{O}}(\epsilon)
+\frac{p^2}{\Lambda^2}{\mathcal{O}}(1+{\mathcal{O}}(\epsilon))
+\frac{p^4}{\Lambda^4}{\mathcal{O}}(1+{\mathcal{O}}(\epsilon))
+\cdots\right\},\\
\text{Re}\left(-\frac{4\pi p^4}{MT_{dd}}\right) &&=\Lambda^5\left
\{-{\mathcal{O}}(1+\tilde{c}_0+{\mathcal{O}}(\epsilon^3))
+\frac{p^2}{\Lambda^2}{\mathcal{O}}(1+{\mathcal{O}}(\epsilon))
+\frac{p^4}{\Lambda^4}({\mathcal{O}}(1+{\mathcal{O}}(\epsilon))-1)
+\cdots\right \}\nonumber\\
&&= \Lambda^5\left \{{\mathcal{O}}(\epsilon)
+\frac{p^2}{\Lambda^2}{\mathcal{O}}(1+{\mathcal{O}}(\epsilon))
+\frac{p^4}{\Lambda^4}({\mathcal{O}}(1+{\mathcal{O}}(\epsilon))-1)
+\cdots\right \}.\eea where a simple fine-tuning for
${\tilde{c}_0}(\equiv \frac{M\Lambda}{4\pi}C_{0;ss})$ is assumed:
${\tilde{c}_0}=-1-{\mathcal{O}}(\epsilon)$, as in Ref.\cite{C71}.
One could also fine-tune $j_0(\equiv\frac{4\pi}{M\Lambda}J_0)$ to
arrive exactly the same effects. Now we see that the simple scenario
could lead us to the following rough estimates for the $^3S_1$
channel: (1) unnatural scattering length, $1/a\sim
\Lambda{\mathcal{O}}(\epsilon)$; (2) natural effective range
$r_e\sim \frac{{\mathcal{O}}(1)}{\Lambda}$, natural $v_2\sim
\frac{{\mathcal{O}}(1)}{\Lambda^3}$, etc. The unnaturalness of the
scattering length lies in the cancelation between $J_0$ and
$\frac{\mathcal{N}_0}{\mathcal{D}_0}|_{p^2=0}$, the same mechanism
as the $^1S_0$ case\cite{C71}. For $^3D_1$ channel, the scattering
length also seems unnatural: $1/a \sim
\Lambda^5{\mathcal{O}}(\epsilon)$, this time it is due to the
cancelation mechanism in the numerator
${\mathcal{N}_0+J_0\mathcal{D}_0}\propto
J_0+\mathcal{N}_0/\mathcal{D}_0$ of the rational term in $1/T_{dd}$,
provided the multiplying factor $\frac{D_0}{N_1+{\mathcal{I}_0}D_1}$
is naturally sized. The rest seem natural, an exception might be
$v_2$ as $J_0$ was pushed into $v_2$ by the factor $p^4$ in ERE in
this channel. In our simple scenario, the main source of
unnaturalness is the physical scale $J_0$ and its universal presence
in all channels\cite{scenario}.

Of course, with different power counting and renormalization
scenarios, we would end up with different magnitude patterns, and
hence different naturalness/unnaturalness in the ERE parameters, or
different naturalness/unnaturalness in scattering behaviors. This is
nothing else but the nontrivial prescription dependence in
nonperturbative regimes. The detailed comparisons across schemes
will be presented later\cite{forth}, where other scenarios like that
devised by KSW\cite{KSW} as well as the original one by
Weinberg\cite{WeinEFT} will be compared. Obviously, such nontrivial
prescription dependence implies that only one set of equivalent
scenarios could faithfully describe the corresponding physics.
Before closing our presentation, we briefly remark on the
differences between our approach and those in literature, especially
the partly perturbative approach of KSW, though we all arrive at
unnatural scattering lengths with fine-tuning: (1) The framework
here is entirely nonperturbative and a new RG invariant scale is
obtained from general analysis in nonperturbative regime, while this
scale is an ordinary running parameter in KSW approach; (2) In
nonperturbative regime, the fine-tuning happens between physical
parameters or quantities at any higher orders: $J_0$ and
$\frac{\mathcal{N}_0}{\mathcal{D}_0}$ that are all RG invariant,
while in KSW, it is a special solution that is only valid at the
lowest order where $1/C_0+J_0$ is physical or RG invariant; (3) As
the KSW power counting differs from that given above, there will be
further different consequences about physical behaviors, which will
be comprehensively compared a later report\cite{forth}. It is also
obvious that, our approach could in principle apply to other
problems that are mainly beset with nonperturbative divergences,
especially those with short range interactions.

{\em Summary.} In this report, we briefly presented the rigorous
solutions of the LSE for $NN$ scattering in the coupled channels
$^3S_1$$-$$^3D_1$ with contact potentials and exposed the main
features of the nonperturbative renormalization of the $T$-matrices
in coupled channels. We also sketched the consequences of a simple
scenario for nonperturbative renormalization proposed before to the
effective range parameters in $^3S_1$$-$$^3D_1$ channels, where an
unnatural scattering length for $^3S_1$ channel could arise with the
rest ERE parameters of $^3S_1$ channel natural. More detailed
results and analysis as well as discussions will be presented in
\cite{forth}. The notions and experiences about nonperturbative
renormalization of $T$-matrices developed here within the context of
contact potentials might be helpful in the treatment of the case
with pion exchanges potentials, as the main divergences involved are
still power like ones and the framework is still in nonperturbative
regime.

{\em Acknowledgement} This project is supported in part by the
National Natural Science Foundation under Grant No.s 10205004 and
10475028 and by the Ministry of Education of China.

\end{document}